\documentclass{iopart}
\usepackage[dvips]{graphicx}
\usepackage[english]{babel}

\newcommand{\be}{\begin{equation}}
\newcommand{\ee}{\end{equation}}

\begin{document}

\title{Bragg spectroscopy of a strongly interacting Bose-Einstein condensate}
\author{J. J. Kinnunen and M. J. Holland}
\address{JILA and Department of Physics, University of Colorado at Boulder, Colorado 80309-0440, USA}
\eads{jjkinnun@colorado.edu}

\begin{abstract}
We study Bragg spectroscopy of a strongly interacting Bose-Einstein
condensate using time-dependent Hartree-Fock-Bogoliubov theory. We include
approximatively the effect of the momentum dependent scattering amplitude
which is shown to be the dominant factor in determining the spectrum 
for large momentum Bragg scattering. The condensation of the
Bragg scattered atoms is shown to 
significantly alter the observed excitation spectrum by creating a novel
pairing channel of mobile pairs.
\end{abstract}

\submitto{\NJP}

\maketitle

\section{Introduction}

A strongly interacting Bose-Einstein condensate (BEC) has been rich
topic to study both theoretically and experimentally. Theoretical problems
arise from the need to provide a proper description of the elementary 
excitations and experimental 
difficulties from the instability of the BEC to three-body collisions.
Research on these systems offers the potential to be highly rewarding 
by shedding light on other strongly interacting systems, such as superfluid
$^4$He.

Weakly interacting BECs, on the other hand, are generally well understood. 
The ground
state properties are accurately described by Bogoliubov theory as the
measurement of the excitation spectrum in 1999 using
Bragg spectroscopy confirmed~\cite{Stenger1999,StamperKurn1999}. In Bragg 
spectroscopy, the condensate is excited using stimulated two-photon 
Bragg scattering yielding the possibility for 
very high momentum and energy resolution. 
The measured spectrum was in very good agreement with the Bogoliubov 
theory~\cite{Steinhauer2002,Tozzo2003}.

The Bragg spectrum of a strongly interacting BEC was measured in 
2008~\cite{Papp2008}. The study found significant deviations from 
the standard Bogoliubov theory, highlighting the need for a more 
thorough theoretical treatment of the process. 

In this work, we study the Bragg specroscopy of a strongly interacting
BEC using Hartree-Fock-Bogoliubov (HFB) 
theory~\cite{Griffin1996}. The theory has been widely applied to various 
problems, both static and time-dependent. In spite of well known problems 
in the excitation spectrum, such as the unphysical zero-momentum gap, it has 
been able to provide correct qualitative 
features and good agreement with experiments~\cite{Holland2001,Kokkelmans2002,Milstein2003,Wuster2005}. 
The present time-dependent theory is an extension of the one used 
in~\cite{Kokkelmans2002}, including both a proper 
renormalization and allowing for the macroscopic occupation 
of the Bragg scattered states but neglecting the explicit molecule 
formation channel. We will show how the appearance of condensate atoms
in finite momentum states leads to profound effects in the
observed Bragg spectrum.

\section{Hamiltonian}

The system is described by the Hamiltonian
\be
\fl   \hat H(t) = \sum_{\vec k} \epsilon_k \hat c_{\vec k}^\dagger \hat c_{\vec k} + \frac{1}{2V} \sum_{\vec k \vec p \vec q} U_{k,p} \hat c_{\vec k + \vec q/2}^\dagger \hat c_{-\vec k+\vec q/2}^\dagger \hat c_{-\vec p+\vec q/2} \hat c_{\vec p + \vec q/2} + \hat H_\mathrm{Bragg} (t),
\label{eq:ham}
\ee
where $\hat H_\mathrm{Bragg} (t)$  is the perturbation due to the Bragg field, $\hat c_{\vec k}$ ($\hat c_{\vec k}^\dagger$) is the annihilation (creation) operator for a boson with momentum $\vec k$, $\epsilon_k = \frac{\hbar^2 k^2}{2m}$, $V$ is the volume, and $U_{k,p}$ is the two-body
T-matrix of the atom-atom interaction. 

As already observed in~\cite{Papp2008}, the momentum dependence of the scattering amplitude is the dominating effect in the high momentum Bragg
scattering and therefore we have included it in the results. A constant interaction strength $U_{k,p} = U$ would correspond to the 
delta function approximation of the real-space T-matrix (yielding constant $U$ in the momentum space). However, we want to include the energy dependence of the two-body T-matrix 
in an approximative way. For low momenta $k$ the energy (or the momentum) 
dependence of the two-body T-matrix for 
on-shell scatterings ($k = p$) is
\be
    U_{k,k} \approx \frac{4\pi\hbar^2a}{m} \Re\left[ \frac{1}{1 + i2ka} \right],
\ee
where $a$ is the scattering length. In order to satisfy the time reversal symmetry of the scattering process (or the Hermiticity of the Hamiltonian~(\ref{eq:ham})), the energy dependent interaction strength $U_{k,p}$ must be symmetric in the exchange of $k$ and $p$. To keep the 
numerical solution tractable, we approximate the interaction
strength by two constant interaction strengths $U_\mathrm{L} := U_{0,0}$ and $U_\mathrm{H} := U_{Q_\mathrm{B},Q_\mathrm{B}}$ so that 
$U_{k,p} = U_\mathrm{L}$ in the low momentum manifold $k,p \ll Q_\mathrm{B}$ and $U_{k,p} = U_\mathrm{H}$ in the high momentum manifold ($k$ and/or $p \approx Q_\mathrm{B}$). That is, the Bragg scattered condensed atoms will interact through interaction strength $U_\mathrm{H}$ whereas the atoms in the unscattered
condensate will feel $U_\mathrm{L}$. 
While the quantum
fluctuations and the Bragg scattering produce excitations with momentum
higher than $Q_\mathrm{B}$, the typical fraction of such atoms is at most $4\%$
in our calculations. Thus, in order to speed up the calculation, we assume
that all fluctuations interact through $U_\mathrm{L}$.

The operator $\hat H_\mathrm{Bragg} (t)$ in the rotating wave approximation can be written as
\be
  \hat H_\mathrm{Bragg} (t) = \Omega(t) \sum_{\vec k} \hat c_{\vec k + \vec Q_\mathrm{B}}^\dagger \hat c_{\vec k} + H.c.,
\ee
where $\vec Q_\mathrm{B}$ is the momentum of the Bragg field and $\Omega(t)$ 
is the coupling between the atoms and the Bragg field. Notice that the 
rotating wave approximation is used here to remove the counter rotating terms that violate the energy conservation. The energy of the Bragg 
field (energy difference between the photons in the two probing laser fields), 
$\hbar \omega_\mathrm{B}$, is included in the coupling $\Omega(t)$.
For a pulse of length $T$ we have 
$\Omega (t) = \Omega \Theta (T-t) e^{i\hbar \omega_\mathrm{B} t}$, where 
$\Omega$ is the strength of the coupling and $\Theta(x)$ is the Heaviside 
function. Using a smooth Gaussian pulse instead of the rapid switching on/off would 
reduce ringing oscillations in many of the results below. However, the step function
is more convenient and straight-forward to implement numerically.

The Hartree-Fock-Bogoliubov theory for BECs has been widely used and its
properties and problems are well known~\cite{Griffin1996,Shi1998}. One 
important problem,
especially related to spectroscopy, is the excitation gap at low
momenta. This is in violation with the Hugenholtz-Pines theorem~\cite{Hugenholtz1959} and
consequently the HFB theory does not yield a proper phonon spectrum. In
practice this means that the Bragg spectrum will have some pathological
features when the Bragg momentum $Q_\mathrm{B}$ is small. In that region, 
we can hope to gain
at most a qualitatively correct picture, assuming that the effect of the
excitation gap is well understood. On the other hand, for large momenta, as
is the case we consider here, the HFB theory is expected to work well.

Another problem in HFB theory relevant to this work is the ultraviolet
divergence of the pairing field $m_0$ when the theory is not correctly 
renormalized. 
This is a consequence of the assumption of a contact
interaction potential which is unable to properly describe high energy
scattering processes. A correct result can still be obtained by employing
a contact interaction
approximation to the effective two-body T-matrix. However, since the
calculation of the many-body T-matrix replicates the same set of diagrams, 
the bare two-body scattering diagrams need to be removed from the 
many-body picture. 
The regularization has been done in several different ways in the literature
but effectively all remove the asymptote of the high-momentum scattering terms.

For a Bose-condensed system, the regularization procedure needs to
be chosen carefully. Indeed, we have found out that the standard
procedure used in several publications~\cite{Giorgini2000,Kokkelmans2002b,Kokkelmans2002,Milstein2003,Wuster2005}
leads into instabilities
by making the energies of low momentum excitations imaginary. This 
regularization issue will be discussed in more detail below.

\section{Equations of motion}

Using the Hamiltonian~(\ref{eq:ham}), we derive the Heisenberg equation 
of motion
\be
\fl  i\hbar \frac{d}{dt} \hat c_{\vec k} = \epsilon_k \hat c_{\vec k} + \sum_{\vec q \vec p} U_{\vec k + \vec q/2,\vec p - \vec q/2} \hat c_{-\vec k + \vec q}^\dagger \hat c_{-\vec p + \vec q} \hat c_{\vec p} + \Omega (t) \hat c_{\vec k - \vec Q_\mathrm{B}} + \Omega(t)^* \hat c_{\vec k + \vec Q_\mathrm{B}}.
\label{eq:eomck}
\ee
For condensed states, $\vec k = l \vec Q_\mathrm{B}$, 
where $l$ is an integer, this operator is 
allowed to have a nonvanishing expectation value 
$\langle \hat c_{n \vec Q_\mathrm{B}} \rangle =: \psi_n$ 
but otherwise the first non-zero terms include 
higher order correlators such as the normal Green's function 
$\langle \hat c_{\vec k}^\dagger \hat c_{\vec k} \rangle$ and the 
anomalous Green's 
function $\langle \hat c_{\vec k} \hat c_{-\vec k} \rangle$.  These
correlators describe both thermal and quantum fluctuations and give corrections
to the Gross-Pitaevskii equation for the condensate. The equation of 
motion~(\ref{eq:eomck}) produces an infinite series of higher order 
correlators. We truncate this series at
the two operator correlator level. 

Due to the Bragg
field, the number of relevant mean-fields is large and the resulting
equation of motion is potentially involved. However, for sufficiently 
weak Bragg
pulses, multiphoton scattering, in which the atom gets kicked 
twice by the Bragg field into momentum state $\pm 2\vec Q_\mathrm{B}$, is 
very unlikely. 
In practice,  we include only the condensate states 
with momenta $-\vec Q_\mathrm{B}$, $0$, and $\vec Q_\mathrm{B}$. 

The equations of motion for the condensates are
\begin{eqnarray}
\nonumber \fl  i\hbar \frac{d}{dt} \psi_0 = \left( \epsilon_0  + h_\mathrm{L} - U_\mathrm{L} \left| \psi_0 \right|^2 \right) \psi_0 + 2 U_\mathrm{H} \left(n_1 \psi_{-1} + n_1^* \psi_1\right) + \left( \Delta_0 - U_\mathrm{L} \psi_0^2 \right) \psi_0^* \\
+ U_\mathrm{H} m_1 \psi_1^* + U_\mathrm{H} m_{-1} \psi_{-1}^* + \Omega (t) \psi_{-1} + \Omega (t)^* \psi_1,
\label{eq:eompsi0mean}
\end{eqnarray}
\begin{eqnarray}
\nonumber \fl  i\hbar \frac{d}{dt} \psi_1 = \left( \epsilon_{Q_\mathrm{B}}  + h_\mathrm{H} - U_\mathrm{H} \left| \psi_1 \right|^2 \right) \psi_1 + 2 U_\mathrm{H} \left( n_1 \psi_0 + n_1^* \psi_2 \right) \\
+ U_\mathrm{H} \left( \psi_0^2 + m_0 \right) \psi_{-1}^* + U_\mathrm{H} m_1 \psi_0^* + \Omega (t) \psi_0,
\label{eq:eompsiQmean}
\end{eqnarray}
\begin{eqnarray}
\nonumber \fl  i\hbar \frac{d}{dt} \psi_{-1} = \left( \epsilon_{Q_\mathrm{B}}  + h_\mathrm{H} - U_\mathrm{H} \left| \psi_{-1} \right|^2 \right) \psi_{-1} + 2U_\mathrm{H} \left( n_1^* \psi_0 + n_1 \psi_2 \right) \\
+ U_\mathrm{H} \left( \psi_0^2 + m_0 \right) \psi_1^* + U_\mathrm{H} m_{-1} \psi_0^* + \Omega (t)^* \psi_0,
\label{eq:eompsiminQmean}
\end{eqnarray}
and for the fluctuations
\begin{eqnarray}
\nonumber \fl  i\hbar \frac{d}{dt} \hat c_{\vec k} = \left( \epsilon_{\vec k} + h_\mathrm{L} \right) \hat c_{\vec k} + 2 U_\mathrm{H} \left( \delta_{Q}^* \hat c_{\vec k + \vec Q_\mathrm{B}} + \delta_{Q} \hat c_{\vec k - \vec Q_\mathrm{B}} \right) +\Delta_0 \hat c_{-\vec k}^\dagger \\
+ \Delta_1 \hat c_{-\vec k + \vec Q_\mathrm{B}}^\dagger + \Delta_{-1} \hat c_{-\vec k - \vec Q_\mathrm{B}}^\dagger 
+ \Omega (t) \hat c_{\vec k - \vec Q_\mathrm{B}} + \Omega (t)^* \hat c_{\vec k + \vec Q_\mathrm{B}}.
\label{eq:eomckmean}
\end{eqnarray}
The Hartree shift of low momentum atoms is $h_\mathrm{L} = 2U_\mathrm{L} \left| \psi_0 \right|^2 + 2U_\mathrm{L} n_0 + 2U_\mathrm{H} \left| \psi_1 \right|^2 + 2U_\mathrm{H} \left| \psi_{-1} \right|^2$, 
where $n_0 = \sum_{\vec k} \langle \hat c_{\vec k}^\dagger \hat c_{\vec k} \rangle$ is the fraction of atoms in the excitations, and $h_\mathrm{H} = 2U_\mathrm{H} n$, where $n$ is the total density of the gas.
The off-diagonal fluctuation density
$n_1 = \sum_{\vec k} \langle \hat c_{\vec k}^\dagger \hat c_{\vec k + \vec Q_\mathrm{B}} \rangle$ 
and $\delta_1 = U_\mathrm{H} \left( \psi_0^\dagger \psi_1 + \psi_{-1}^\dagger \psi_0 + n_1 \right)$. The mean-field pairing fields are defined as 
$\Delta_0 = \left( U_\mathrm{L} \psi_0 \psi_0 + U_\mathrm{H} 2\psi_1 \psi_{-1} + U_\mathrm{L} m_0 \right)$ and
$\Delta_{\pm 1} = U_\mathrm{H} \left( 2\psi_0 \psi_{\pm 1} + m_{\pm 1} \right)$, where 
$m_n = \sum_{\vec k} \langle \hat c_{\vec k} \hat c_{-\vec k + n\vec Q_\mathrm{B}} \rangle$.
Notice that the first three equations of motion (\ref{eq:eompsi0mean}), (\ref{eq:eompsiQmean}), and (\ref{eq:eompsiminQmean})
are mean-field condensates but the last one~(\ref{eq:eomckmean}) is an equation of motion for a
fluctuation operator. From the fluctuation operator we form equations of
motion for the fluctuation fields $\langle \hat c_{\vec k}^\dagger \hat c_{\vec p} \rangle$ and $\langle \hat c_{\vec k} \hat c_{-\vec p} \rangle$. 
All anomalous fluctuation fields
$m_0$ and $m_{\pm 1}$ are ultraviolet divergent and need to be
regularized.

Instead of coupling the condensate $\psi_0$ directly into the excitations 
$\langle \hat c_k^\dagger \hat c_k \rangle$, the Bragg field $\Omega$ provides
the coupling into mobile condensate states $\psi_{\pm 1}$. Initially these 
mobile
condensate states are empty but the Bragg pulse will break the translational
symmetry of the condensate by rotating the condensed atoms into a superposition
of the zero-momentum condensate state and the mobile condensate states.
The two-body
scattering processes couple these atoms into excitations, 
eventually leading into 
dephasing of the single-particle density matrix and, in principle, to
fragmentation into separate condensates $\psi_0$ and $\psi_{\pm 1}$.  
As will be shown below, this coupling
is relatively strong as it leads into rapid decay of the excited condensates.
However, the present theory is unable to describe the dephasing process.

The set of equations of motion above shows that the Bragg field creates also
new excitation fields, such as $\langle \hat c_{\vec k}^\dagger \hat c_{\vec k + \vec Q_\mathrm{B}} \rangle$ and 
$\langle \hat c_{\vec k} \hat c_{-\vec k \pm \vec Q_\mathrm{B}} \rangle$. In the 
calculations below, we have included the equations of motion for all these 
fields in addition to the standard excitation fields 
$\langle \hat c_{\vec k}^\dagger \hat c_{\vec k} \rangle$ and 
$\langle \hat c_{\vec k} \hat c_{-\vec k} \rangle$. It is these mean-fields and the
condensate fields $\psi_0$, $\psi_{\pm 1}$ that we propagate in real time in 
our theory. As our 
analysis below shows, all
these fields are needed for a full description of even relatively weak
Bragg pulses in which only a small fraction of atoms is excited by the field.
In particular, the backward scattered condensate $\psi_{-1}$ is important in 
order to obtain the proper Bogoliubov spectrum in the weakly interacting limit.

In all the numerical calculations below, we have considered a uniform $^{85}$Rb 
condensate of density $10^{14}\,\mathrm{cm}^{-3}$.

\section{Regularization of the ultraviolet divergence}

Before studying the Bragg spectroscopy any further, we will address some
issues related to the regularization of the anomalous fluctuation fields. 
Using Matsubara Green's functions, the anomalous fluctuation field $m_0$
can be written as
\be
   m_0 = \frac{1}{\beta} \sum_{\vec K} G ({\vec K}) G_0(-{\vec K}),
\ee
where $G_0({\vec K})$ is the bare Bose Green's function and $G({\vec K})$
is the dressed Green's function obtained for some self-energy $\Sigma$. The
four-vector ${\vec K} = (i\omega,{\vec k})$ consists of a 
Matsubara frequency $i\omega$ and a three-dimensional momentum $\vec k$. 
The regular Hartree-Fock-Bogoliubov theory gives, and is given by, 
\be
   G(\vec K) = u_k^2 \frac{1}{i\omega - E_k} - v_k^2 \frac{1}{i\omega + E_k},
\ee
where $u_k^2,v_k^2 = \frac{1}{2} \left( \frac{\epsilon'_k}{E_k} \pm 1 \right)$,
$\epsilon'_k = \epsilon_k + 2Un - \mu$, $E_k = \sqrt{\epsilon_k^{'2} - \Delta^2}$, and $\Delta = U\left|\psi_0\right|^2 + Um_0$. However, the important point
is that the bare Green's function needs to include the Hartree shifts, i.e.
\be
  G_0(\vec K) = \frac{1}{i\omega - \epsilon'_k}.
\label{eq:baregreen}
\ee
Regularization can now be done self-consistently by removing the free particle
diagrams where the free particle propagator is given by the same energy shifted
Green's function $G_0$. Thus we remove from $m_0$ the term
\be
   m_0^\mathrm{2B} = \frac{1}{\beta} \sum_{\vec K} G_0({\vec K}) G_0(-{\vec K}).
\ee

This regularization differs from the standard scheme used in
the literature~\cite{Giorgini2000,Kokkelmans2002b,Kokkelmans2002,Milstein2003,Wuster2005} by the energy shift $2Un - \mu$ in the bare Green's function $G_0({\vec K})$.
Neglecting the energy shift, i.e. replacing $\epsilon'_k$ by $\epsilon_k$ in
the regularizing term $m_0^\mathrm{2B}$ will make the regularizing part
larger than the initial anomalous term $\left|m_0 \right| < \left| m_0^\mathrm{2B} \right|$. This has the unfortunate effect of turning the regularized
anomalous pairing field positive. The low momentum energy gap in the
HFB spectrum $\sqrt{-4Un_0 m_0}$ becomes then imaginary and the low energy
excitations turn unstable. However, the imaginary energies are small and the
corresponding lifetimes are long, so that the instability can easily go 
unnoticed. In addition, instabilities apply only to the very lowest energy 
excitations requiring a dense grid in the momentum space in order to play a 
role. 

Including a constant energy shift $\delta$ in the bare 
Green's function does not affect the two-body scatterings but guarantees 
$m_0 < 0$ if $\delta$ is large enough. Indeed, choosing $\delta = 2Un - \mu$, 
as in ~(\ref{eq:baregreen}),
is enough to guarantee positivity of $m_0$. However, this would require
an iterative solution of $\delta$ as the chemical potential $\mu$ depends
on $\delta$. Here we approximate $\delta = Un$ but the results are relatively 
insensitive to the actual choice of $\delta$ as long as it is large enough. 
Regularization of the mobile
pairing fields $m_{\pm 1}$ proceeds in an identical manner.

\section{Bragg spectroscopy of a weakly interacting BEC}

We will first study the Bragg spectroscopy of a weakly interacting Bose gas.
Since the vast majority of the atoms are in the initial condensate state $\psi_0$, 
we are interested only in the energy of the excited condensate state with 
momentum
$\vec Q_\mathrm{B}$. In the weakly interacting regime, we can safely ignore 
the fluctuation parts of the finite momentum pairing and density fields 
$m_{\pm 1}$ and $n_1$.  The equations of motion of the excited 
condensates are now (assuming that the populations of the excited condensate
states $\psi_1$ and $\psi_{-1}$ are small)
\be
  i\hbar \frac{d}{dt} \psi_1 = \left( \epsilon_{Q_\mathrm{B}} + h_\mathrm{H} \right) \psi_1 + U_\mathrm{H} \left( \psi_0^2 + m_0 \right) \psi_{-1}^* + \Omega(t) \psi_0
\label{eq:eompsiQweak}
\ee
and
\be
  i\hbar \frac{d}{dt} \psi_{-1} = \left( \epsilon_{Q_\mathrm{B}} + h_\mathrm{H} \right) \psi_{-1} + U_\mathrm{H} \left( \psi_0^2 + m_0 \right) \psi_1^* + \Omega (t)^* \psi_0.
\label{eq:eompsiminQweak}
\ee
In the absence of the Bragg coupling $\Omega (t) = 0$, 
these equations can be solved by the Bogoliubov transformation, showing
that the energies of the excited condensates at momenta $\pm\vec Q_\mathrm{B}$
do indeed match the energies of the corresponding excitations 
in the Hartree-Fock-Bogoliubov specturm.

Notice that, in addition to the Bragg field, the zero-momentum pairing field 
$\Delta_0$ acts as a source for the two condensates $\psi_{\pm 1}$ even though 
it cannot level out the difference in the populations due to momentum 
conservation (i.e. a zero momentum pair of 
atoms can be turned into a pair $\psi_1 \psi_{-1}$ which 
conserves momentum but also the population difference in the two states).
This stability of the population difference between the $\psi_1$ and 
$\psi_{-1}$ condensates underlines the importance of the off-diagonal
fields $\Delta_{\pm 1}$ and $n_1$ for the decay of the multiply condensed state.
Indeed, it is only through these fields that the system finally decays
into an equilibrium state. However, this decay process is slow in a weakly
interacting gas, and can therefore be neglected in most cases.

\section{Bragg spectroscopy of a strongly interacting BEC}

For strong interactions, the off-diagonal fields are needed for the decay 
processes and for the correct excitation spectrum. The presence of 
additional pairing fields
$\Delta_{\pm 1}$ allows the atoms to lower their energies even further 
by pair formation and this is reflected also in the Bragg spectrum. 
\begin{figure}
\includegraphics[height=7cm]{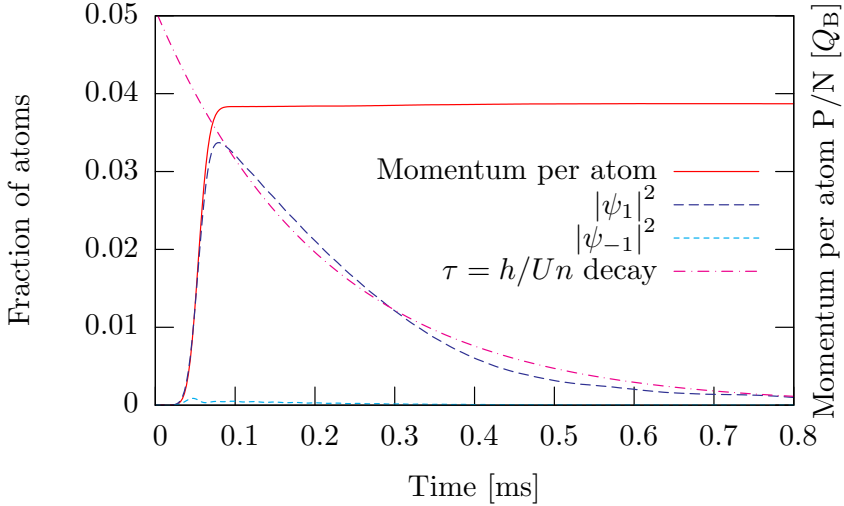}
\caption{Evolution of the excited condensates as a function of time. The atoms in
the excited condensates decay into ordinary excitations following roughly an 
exponential decay with lifetime of $h/Un$. After the Bragg pulse
(of length $0.1\,\mathrm{ms}$) the total momentum $P$ is conserved,
showing that the momentum of the excited condensates is indeed transferred
to the excitations. Here $na^3 \approx 0.003$.}
\label{fig:lifetime}
\end{figure}
Fig.~\ref{fig:lifetime} shows the decay of the excited condensates after (and during)
the Bragg pulse. The decay lifetime scales as $h/Un$, showing that the effect of
the mobile pairing field is large even though the anomalous fields $m_{\pm 1}$ are
small. Notice that the total momentum is not quite conserved after the pulse. This is due to
a finite momentum cutoff in the summations, but the error can be made arbitrarily small
by increasing the momentum grid size. 

\begin{figure}
\includegraphics[height=7cm]{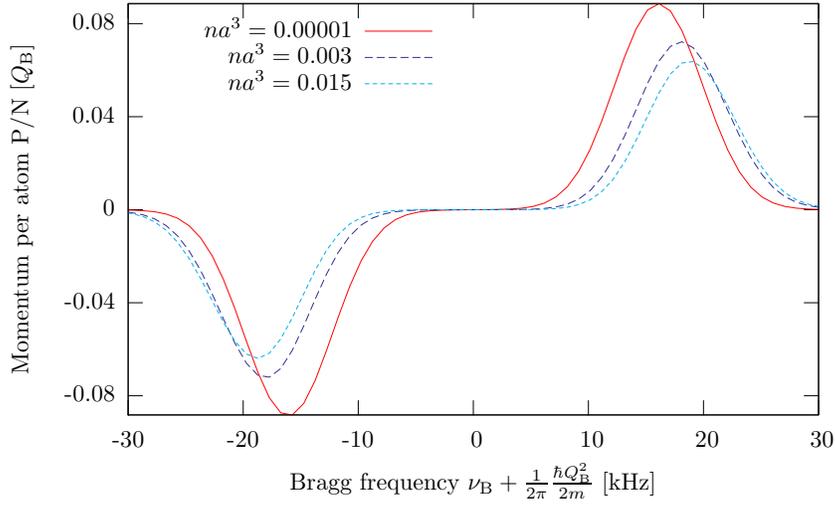}
\caption{The total momentum per atom $P/N$ given by the Bragg pulse as a function of Bragg
detuning energy ($h \nu_\mathrm{B} = \hbar \omega_\mathrm{B} - \frac{\hbar^2 Q_\mathrm{B}^2}{2m}$) for three different interaction strengths $na^3 = 10^{-5}$, $na^3 = 0.003$, and $na^3 = 0.015$. 
For very short or strong Bragg pulses the two peaks
overlap. The Bragg resonance can be resolved only when the two peaks are
separate. 
The free particle resonance $\frac{\hbar^2 Q_\mathrm{B}^2}{2m} \approx h 15.4\,\mathrm{kHz}$. 
} 
\label{fig:spectrum}
\end{figure}
Fig.~\ref{fig:spectrum} shows the Bragg spectrum for a Bragg pulse exciting 
roughly $10\%$ of atoms. Here and in the rest of the figures Bragg 
detuning $h \nu_\mathrm{B}$ refers to 
the Bragg field energy offset from the free particle resonance 
$h \nu = \hbar \omega_\mathrm{B} - \frac{\hbar^2 Q_\mathrm{B}^2}{2m}$.
The figure shows two peaks corresponding to the forward and
backward scattering and the shift from the free particle line with increasing
interaction strength. The line shape also changes from a Lorentzian (for weakly
interacting gas) into an asymmetric peak (for strongly interacting gas) because
also quantum fluctuations are affected by the Bragg field and the corresponding atoms
have the Bragg resonance at higher detuning than the condensate atoms (because
the excitation spectrum is a concave function of momentum $k$). 
For finite temperatures, the asymmetricity of the peak would be even
more pronounced as the condensate fraction is reduced.

\begin{figure}
\includegraphics[height=7cm]{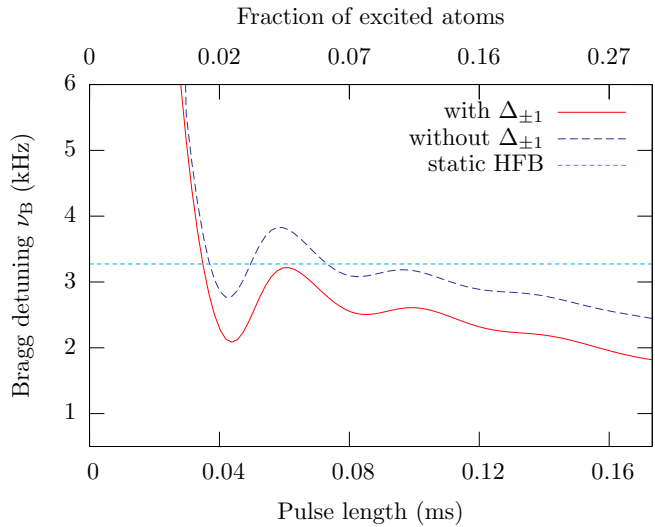}
\caption{The position of the Bragg spectrum maximum as a function of the
Bragg pulse length.
For very short pulses, the lifetime broadening mixes the backward and
forward scattering peaks. For slightly longer pulses the peaks separate and
the resonance energy can be resolved. The gradual drift in the peak position
for long pulses is caused by the depletion of the initial condensate. 
Here the interaction
strength is $na^3 = 0.003$ and shown are plots both with and without the 
mobile pairing fields $\Delta_{\pm 1}$, and the static Hartree-Fock-Bogoliubov
result (the straight horizontal line). The top x-axis shows the approximate
fractions of excited atoms for the corresponding pulse length.}
\label{fig:peakpos}
\end{figure}
From spectra such as shown in Fig.~\ref{fig:spectrum} we determine the 
position of the peak maximum and the Fig.~\ref{fig:peakpos} 
shows how this 
evolves as a function of pulse length. For very short pulses the 
linewidth is very large due to the lifetime broadening and the
two peaks in the spectra overlap. Thus, the position of the peak maximum 
depends strongly on the pulse length. Once the pulse is long enough, so that
the linewidth of the pulse is less than the distance between the backward 
and forward scattering peaks, the Bragg resonance can be resolved. There 
is still some slow drift in the peak position due to the depletion of the
initial condensate, causing some error in the determination of the resonance
energy. As long as the total fraction of excited atoms is small enough (less than $10\%$), the Bragg spectrum is insensitive to the strength of the Bragg 
coupling. Therefore, in order to be able to resolve the Bragg resonance well, 
one can use very weak pulses. Furthermore, a Gaussian Bragg pulse would reduce
the initial oscillations in the spectrum peak position. 
In an actual experimental setup, the maximum
length of the pulse is limited due to the finite size of the system and the 
inhomogenous trapping potential. The Fig.~\ref{fig:peakpos} shows also how 
well the time-dependent Hartree-Fock-Bogoliubov theory reproduces the 
resonance energy of the corresponding static theory when the mobile pairing
fields are neglected.

\begin{figure}
\includegraphics[height=7cm]{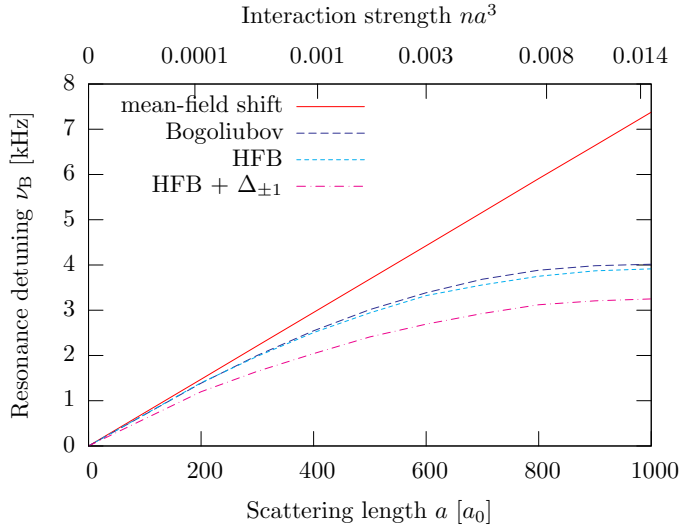}
\caption{The Bragg resonance detuning $\nu_\mathrm{B}$ as a function of the interaction
strength for various theories. Shown are mean-field shift spectrum $Un$, 
Bogoliubov spectrum, Hartree-Fock-Bogoliubov spectrum, and the 
spectrum for HFB theory including mobile pairing fields $\Delta_{\pm 1}$.  
Bogoliubov line is roughly the same as the $\sqrt{8\pi na^3}$ line
in Fig.1 of~\cite{Papp2008} (the densities in the two plots are different).}
\label{fig:soup}
\end{figure}
Fig.~\ref{fig:soup} shows the Bragg resonance energy as a function of the 
interaction strength. At strong interactions, the resonance drops far from
the mean-field shift line. Notice that the Bogoliubov line includes 
the effect of $k$-dependent scattering length and it agrees surprisingly
well with the HFB theory. Without the $k$-dependent scattering length, the
Bogoliubov line would be close to the mean-field shift line.
The mean-field shift and the Bogoliubov line are calculated using a 
static theory 
but the Hartree-Fock-Bogoliubov lines are from the present time-dependent 
theory. The presence of the mobile pairing fields drops the resonance energy
even further.

Neglecting the effect due to the mobile pairing fields $\Delta_{\pm 1}$, the 
effect of the $na^3$ corrections introduced by the HFB theory (as compared to 
the Bogoliubov result) is very small despite much larger effect on the
chemical potential (roughly $10\%$ for $na^3=0.015$). Interestingly,
in the $na^3$ expansion of the corrections to the Bogoliubov theory, the
leading order term in the chemical potential 
(the Lee-Huang-Yang (LHY) correction~\cite{Lee1957, Lee1957b}) is $74\%$. 
However, 
the next higher order correction~\cite{Wu1959} is
roughly $150\%$ showing that the standard $na^3$ expansion is breaking
down.
It is also interesting to notice that the Popov approximation 
of the HFB theory (HFB-Popov theory) does reproduce
correctly the LHY results in the leading order of the $na^3$ expansion. 
However, there are also higher order terms included.
The HFB theory does actually 
agree very well with the gapless HFB-Popov theory for large Bragg momenta,
and therefore we expect that the LHY terms are properly included in the 
present theory.
The different treatment of the beyond LHY terms may be the reason
for the discrepancy between our HFB results and the Beliaev 
theory~\cite{Beliaev1958} based results in~\cite{Papp2008}. 

The qualitative features of Fig.~\ref{fig:soup} agree with the experimental 
data in~\cite{Papp2008}. However, since the present study has been 
done for a uniform gas, this work should be extended to a trapped gas 
using local density approximation for proper comparison with the 
experimental results.

\section{Summary}

To summarize, we have studied the Bragg spectroscopy of a strongly interacting
BEC using time-dependent Hartree-Fock-Bogoliubov theory. Taking into account
momentum dependence of the scattering amplitude, the theory is in qualitative
agreement with the experiment done on strongly interacting Rb-85 condensate. 
The most surprising effect comes from the creation of mobile pairing fields
and the subsequent change in the excitation spectrum. While the present
experimental results cannot confirm this effect, it should be more visible
in the low-momentum excitation spectrum. 
Another interesting question would be the relation to fragmented condensates~\cite{Mueller2006}.
Because of the coherence of the Bragg field, the different condensates at momenta $n\vec Q_\mathrm{B}$
have a well defined phase-difference. The coupling to the excited states should,
in principle, dephase the condensate, leading into a fragmented state. 
The Bragg scattering induced mobile pairing fields have also an interesting 
connection to the FFLO-type pairing~\cite{Fulde1964,Larkin1965}
in polarized Fermi gases that would be worth further research.

\ack

This project was supported by U.S. Department of Energy, Office of Basic 
Energy Sciences via the Chemical Sciences, Geosciences, and Biosciences 
Division. We acknowledge useful discussions with S. Ronen, J. Wachter, D. Meiser, and the Rb-85 team in JILA.

\section*{References}
\bibliographystyle{unsrt}
\bibliography{braggBEC}

\end{document}